# Reconstruction and Normalization of Anselin's Local Indicators of Spatial Association (LISA)

Yanguang Chen

(Department of Geography, College of Urban and Environmental Sciences, Peking University, 100871, Beijing, China. Email: chenyg@pku.edu.cn)

**Abstract**: The local indicators of spatial association (LISA) are significant measures for spatial autocorrelation analysis. However, there is an inadvertent fault in Anselin's mathematical processes so that the local Moran and Geary indicators do not satisfy his second basic requirement, i.e., *the sum of the local indicators is proportional to a global indicator*. Based on Anselin's original intention, this paper is devoted to reconstructing the calculation formulae of the local Moran indexes and Geary coefficients through mathematical derivation and empirical evidence. Two sets of LISAs were clarified by mathematical reasoning. One set of LISAs is based on no normalized weights and centralized variable (MI1 and GC1), and the other set is based on row normalized weights and standardized variable (MI2 and GC2). The results show that the first set of LISAs satisfy Anselin's second requirement, but the second the set cannot. Then, the third set of LISA was proposed, treated as canonical forms (MI3 and GC3). The local Moran indexes are based on global normalized weights and standardized variable based on population standard deviation, while the local Geary coefficients are based on global normalized weights and standardized variable based on sample standard deviation. This set of LISAs satisfies the second requirement of Anselin's. The observational data of city population and traffic mileage in Beijing-Tianjin-Hebei region of China were employed to verify the theoretical results. This study helps to clarify the misunderstandings about LISAs in the field of geospatial analysis.

# 1 Introduction

Geography has two core concepts: difference and dependence. The former is related to a classical topic of geography, while the latter is related to spatial correlation analysis. The concept of spatial difference is also termed regional differences, which came from areal differentiation (Hartshorne, 1959; Hu *et al*, 2018; Martin, 2005). The traditional concept of difference seems to be in contradiction with the pursuit of general laws, so geography embarks on the road of "exceptionalism" (Schaefer, 1953). After the quantitative revolution (1953-1976), geography began to attach importance to spatial correlation, which indicates spatial dependence. Gravity models, spatial interaction models, and spatial autocorrelation analysis are the main approaches to research spatial correlation processes (Griffith, 2003; Haggett *et al*, 1977). Spatial autocorrelation is originally a biological statistic concept, which is mainly used to evaluate whether the spatial sampling results meet the traditional statistical requirements (Moran, 1948; Moran, 1950; Geary, 1954). When geographers introduced spatial autocorrelation measure into geospatial analysis, they found that there are few spatial uncorrelated phenomena. In this context, the spatial autocorrelation analysis method was developed (Cliff and Ord, 1973; Cliff and Ord, 1981; Odland, 1988). The early spatial autocorrelation analysis was only at the global level, rarely involving the local level, so it provided limited geospatial information. In other words, the initial spatial autocorrelation focuses on spatial dependence rather than spatial difference. After the theoretical revolution in the later period of the quantitative revolution was frustrated, the traditional regional trend of thought of geography returned quietly, and the concept of regional difference was again valued by geographers with a new expression of spatial heterogeneity (Anselin, 1996). Tobler (1970) proposed the first law of geography based on spatial dependence, and Harvey proposed that spatial heterogeneity be the second law of geography (Tobler, 2004). The study of spatial heterogeneity naturally involves spatial locality. According to Fotheringham (1997, 1998, 1999), there are three trends in the development of quantitative geography: localization, computation and visualization. In this context, local spatial autocorrelation analysis came into being (Anselin, 1995; Anselin, 1996; Getis and Aldstadt, 2004; Getis and Ord, 1992; Ord and Getis, 1995). Therefore, spatial difference (heterogeneity) and spatial correlation (dependency) have reached the same goal through different routes (Anselin, 1996; Goodchild, 2004).

Local spatial autocorrelation analysis is developed on the basis of global spatial autocorrelation analysis. The Local Indicators of Spatial Association (LISA) proposed by Anselin (1995) plays an important role in the local correlation analysis of geographical research. LISA includes local Moran indexes and local Geary coefficients. These spatial statistics, together with the *G* index proposed by Getis and Ord (1992) and Moran scatterplot proposed by Anselin (1996), have become systematic tools for local autocorrelation analysis. However, even the wise are not always free from error. Anselin (1995) made an unintentional mistake in the process of reasoning, which caused some cognitive confusion in geospatial analysis. Two points need to be clarified. Firstly, spatial statistics represent a kind of measures, which may be used to describe or infer. No matter where the goal is, a good measure should have a clear critical value or boundary value (Chen, 2017). For example, the boundary values of Pearson correlation coefficient is -1 and 1, and the critical value is 0. Secondly, if two measures are equivalent to one another, the ratio of the two measures is constant. For example, the ratio of Student's *t* statistic to Pearson's part correlation coefficient is constant, which equals the square root of the ratio of residuals mean square deviation to total sum of squares. Anselin's LISA has two shortcomings: one is the lack of clear boundary value and critical value; the other is that the two sets of local Moran index are not equivalent to each other, and the two sets of local Geary coefficients are not equivalent to each other. One of the key reasons lies in that symmetric spatial contiguity matrix is replaced by asymmetric row normalized spatial weight matrix in the process of mathematical deduction. In addition, the definition of local Geary coefficient is based on the population standard deviation instead of the sample standard deviation, which is not consistent with original aim of defining Geary's coefficient. The purpose of this paper is to sort out Anselin's mathematical reasoning process and correct his unintentional mistakes. Based on the mathematical derivation, the local Moran index and local Geary coefficient will be normalized. Finally, the strict mathematical relationship between Moran's indexes and Geary's coefficients are derived. The observational data of the system of cities in Beijing-Tianjin-Hebei region in China will be employed to testify the improved results.

# 2 Theoretical results

## 2.1 Anselin's spatial autocorrelation measurements

### 2.1.1 Anselin's first formula of local Moran index

One of the bases of spatial analysis is spatial proximity matrix, which can be measured by spatial distance matrix. Spatial distance matrix or spatial proximity matrix can be transformed into spatial contiguity matrix by means of spatial weight function such as negative power law or step function (Chen, 2012; Getis, 2009). Suppose that there are $n$ elements in a geographical region, and this size of the $i$th element is measured by $x_i$ ($i$=1,2,…,$n$). The size variable $x$ are not standardized and the spatial contiguity matrix $\mathbf{V}=[v_{ij}]$ is not transformed into the globally normalized spatial weight matrix $\mathbf{W}=[w_{ij}]$. Using the symbol systems defined in this work, we can extract two sets of local spatial autocorrelation statistics (Table 1). The first local Moran index formula defined by Anselin (1995) is as follows

$$I_i^* = (x_i - \bar{x})\sum_{j=1}^{n} v_{ij}(x_j - \bar{x}) = y_i \sum_{j=1}^{n} v_{ij} y_j, \tag{1}$$

where $y_i = x_i - \bar{x}$, $y_j = x_j - \bar{x}$ denote centralized size variables, and $\bar{x}$ refers to mean value. The centralized variables can be transformed into standardized variables by means of $z$-score formula. Based on population standard derivation, the standardized variables can be expressed as

$$z_i = \frac{y_i}{\sigma} = \frac{x_i - \bar{x}}{\sigma}, \quad z_j = \frac{y_j}{\sigma} = \frac{x_j - \bar{x}}{\sigma},$$

where $z$ denote standardize variable. The sum of equation (1) is

$$\sum_{i=1}^{n} I_i^* = \sum_{i=1}^{n} y_i \sum_{j=1}^{n} v_{ij} y_j = \sum_{i=1}^{n} \sum_{j=1}^{n} v_{ij} y_i y_j, \tag{2}$$

which is essentially the sum of squares of spatial weighted deviations. The sum of the elements in spatial contiguity matrix is

$$V_0 = \sum_{i=1}^{n} \sum_{j=1}^{n} v_{ij}. \tag{3}$$

Dividing equation (1) by $V_0$ yields spatial weighted auto-covariance as follows

$$Cov = \frac{1}{\sum_{i=1}^{n}\sum_{j=1}^{n}v_{ij}}\sum_{i=1}^{n}I_i^* = \frac{1}{V_0}\sum_{i=1}^{n}\sum_{j=1}^{n}v_{ij}y_iy_j. \tag{4}$$

Furthermore, the spatial weighted covariance can be divided by the population variance of the size variable, which is called the second moment by Anselin (1995), that is

$$\sigma^2 = \frac{1}{n}\sum_{i=1}^{n}(x_i - \bar{x})^2 = \frac{1}{n}\sum_{i=1}^{n}y_i^2. \tag{5}$$

The result is global Moran's index, $I=Cov/\sigma^2$. It can be expanded as

$$I = \frac{\frac{1}{\sum_{i=1}^{n}\sum_{j=1}^{n}v_{ij}}\sum_{i=1}^{n}I_i^*}{\frac{1}{n}\sum_{i=1}^{n}y_i^2} = \frac{n\sum_{i=1}^{n}\sum_{j=1}^{n}v_{ij}y_iy_j}{V_0\sum_{i=1}^{n}y_i^2} = \frac{1}{\sigma^2 V_0}\sum_{i=1}^{n}\sum_{j=1}^{n}v_{ij}y_iy_j = \sum_{i=1}^{n}\sum_{j=1}^{n}w_{ij}z_iz_j, \tag{6}$$

where $w_{ij}$ is the element of the global normalized weight matrix **W**. According to Anselin (1995), equation (6) can be expressed as

$$I = \frac{1}{\sigma^2 V_0}\sum_{i=1}^{n}I_i^*. \tag{7}$$

The relationship between the sum of Anselin's first local Moran index and the global Moran index is obtained as below

$$\sum_{i=1}^{n}I_i^* = \sigma^2 V_0 I = \gamma I. \tag{8}$$

The proportionality coefficient in equation (8) is

$$\gamma = \sigma^2 V_0 = (\frac{1}{n}\sum_{i=1}^{n}y_i^2)(\sum_{i=1}^{n}\sum_{j=1}^{n}v_{ij}). \tag{9}$$

Equation (3) can be replaced by a vector indicating the sum of rows of the spatial contiguity matrix as below

$$V_i = \sum_{j=1}^{n}v_{ij}. \tag{10}$$

Spatial contiguity matrix can be normalized by row. Anselin (1995) called it row-standardized spatial weights matrix. In this way, equation (4) becomes a locally weighted spatial auto-covariance, that is

$$Cov_i = \frac{I_i^*}{\sum_{j=1}^{n} v_{ij}} = \frac{1}{V_i} \sum_{j=1}^{n} v_{ij} y_i y_j. \tag{11}$$

The summation of equation (11) is

$$\sum_{i=1}^{n} Cov_i = \sum_{i=1}^{n} \frac{I_i^*}{V_i} = \sum_{i=1}^{n} \sum_{j=1}^{n} \frac{v_{ij}}{V_i} y_i y_j. \tag{12}$$

In this case, it is impossible to obtain the global spatial weighted auto-covariance, and it is impossible to derive the simple summation relationship between local Moran index and global Moran index. If so, the reasoning from equation (4) to equation (9) will be invalid.

It can be seen that the local-global relationship based on Anselin's first local Moran index formula is a global normalized weight matrix with symmetry. The first local Moran index formula of Anselin (1995) is correct, it satisfy the two requirements defined by Anselin (1995). The shortcoming lies in that it is not standardized. A good measure should have a clear critical value (reference value) or a pair of explicit boundary values. However, the local Moran index calculated by equation (1) has neither boundary values nor clear threshold value.

Table 1 Three sets of LISAs researched in this paper based on Anselin's work

| Type | Index | Weight matrix | Size variable | Symbol |
|---|---|---|---|---|
| **First set of local LISA** | Local Moran's $I$ | No normalization | Centralization | MI1 |
|  | Local Geary's $C$ | No normalization | Centralization | GC1 |
| **Second set of local LISA** | Local Moran's $I$ | Row normalization | Standardization based on population standard deviation | MI2 |
|  | Local Geary's $C$ | Row normalization | Standardization based on population standard deviation | GC2 |
| **Third set of local LISA** | Local Moran's $I$ | Global normalization | Standardization based on population standard deviation | MI3 |
|  | Local Geary's $C$ | Global normalization | Standardization based on sample standard deviation | GC3 |

## 2.1.2 Anselin's second formula of local Moran index

Suppose that the variables are standardized, the spatial contiguity matrix is transformed into a spatial weight matrix which is normalized by row. In this way, $V_0$ in is replaced by $V_i$ in equation (4). Thus, revised equation (4) divided by population variance yields the second local Moran's index formula of Anselin (1995), $I_i^{**}=Cov_i/\sigma^2$, that is

$$I_i^{**} = \frac{1}{\sigma^2}\sum_{j=1}^{n}\frac{v_{ij}}{V_i}y_i y_j = \frac{1}{\sigma^2}y_i\sum_{j=1}^{n}w_{ij}^{*}y_j, \quad (13)$$

where $w_{ij}^{*}$ denotes the elements in the row normalized spatial weight matrix, $\mathbf{V}^{*}$. Thus, the sum of the spatial weight matrix is

$$V_0^{*} = \sum_{i=1}^{n}\sum_{j=1}^{n}\frac{v_{ij}}{V_i} = \sum_{i=1}^{n}(1) = n. \quad (14)$$

The variance of standardized variable is 1, namely, $\sigma^2=1$. For normalized matrix by row, the sum is $V_0^{*}=n$, thus we have

$$\gamma = \sigma^2 V_0^{*} = V_0^{*} = n. \quad (15)$$

Substituting equation (15) into equation (8) seems to yield the following relation

$$\sum_{i=1}^{n}I_i^{**} = nI. \quad (16)$$

On the surface, there is no problem at all. The two asterisks indicate the inherent difference between the two sets of local Moran's indexes. However, Anselin (1995) inadvertently made a mistake in above reasoning process.

Mathematical deduction problems can be revealed through logical analysis, and also can be reflected through empirical analysis. Let us check the problem from another view of angle. The relation between the second set of local Moran's indexes of Anselin (1995) and global Moran's index can be derived from equation (13). The summation of the local Moran's indexes based on equation (13) is

$$\sum_{i=1}^{n}I_i^{**} = \frac{1}{\sigma^2}\sum_{i=1}^{n}\sum_{j=1}^{n}\frac{v_{ij}}{V_i}y_i y_j = V_0\sum_{i=1}^{n}\sum_{j=1}^{n}\frac{w_{ij}}{V_i}z_i z_j = \sum_{i=1}^{n}\sum_{j=1}^{n}w_{ij}^{*}z_i z_j. \quad (17)$$

By variable standardization, the population standard deviation becomes 1 unit, i.e., $\sigma^2=1$. However, the row sum of spatial contiguity matrix $V_i$ is not a constant. It can neither be eliminated nor converted to a constant. Therefore, no constant proportionality relation between the second set of local Moran's index and the global Moran's index. If and only if equation (6) is introduced into equation (17) can the proportional relationship similar to equation (8) be derived. Based on equation

(6), equation (17) can be re-expressed as

$$\sum_{i=1}^{n} I_i^{**} = \frac{\sum_{i=1}^{n}\sum_{j=1}^{n} w_{ij}^* z_i z_j}{\sum_{i=1}^{n}\sum_{j=1}^{n} w_{ij} z_i z_j} I . \tag{18}$$

Unfortunately, we cannot prove the following relation:

$$\sum_{i=1}^{n}\sum_{j=1}^{n} w_{ij}^* z_i z_j = n \sum_{i=1}^{n}\sum_{j=1}^{n} w_{ij} z_i z_j = nI . \tag{19}$$

This lend further support to the judgment that equation (16) does not hold. However, the proportional relationship given in equation (18) can be easily verified by the observation data. Another view of angle is to examine the ratios of two sets of local Moran indices. If the ratios are constant, the two definitions are equivalent to one another, otherwise they are not. In fact, the values in the first set of local Moran indexes divided by the corresponding values in the second set of local Moran indexes yields

$$\frac{I_i^*}{I_i^{**}} = \frac{\sigma^2 \sum_{j=1}^{n} v_{ij} y_i y_j}{\sum_{j=1}^{n} \frac{v_{ij}}{V_i} y_i y_j} = \sigma^2 V_i , \tag{20}$$

which, obviously, is a variable that changes with $V_i$ rather than a constant.

It can be seen that the ratios of two sets of local Moran's indexes are not constant, so they are not equivalent to each other. This suggests that, the second set of local Moran indexes cannot satisfy the second requirement of Anselin (1995), which said, "The sum of the local indicators is proportional to a global indicator". The reason for the fault is that Anselin (1995) inadvertently replaced a concept in this mathematical derivation. Concretely speaking, the global normalized symmetric weight matrix **W** becomes the local normalized asymmetric weight matrix **V**$^*$. This way violates the law of identity of concepts and the principle of logical consistency in mathematical reasoning.

### 2.1.3 Anselin's formula of local Geary coefficient

The global Geary coefficient is complementary to the global Moran index: the former is oriented to sample analysis, and the latter is based on statistical population. Similar to the treatment of local Moran index, two local Geary statistics were defined by Anselin (1995). It is assumed that the variables are not standardized and the spatial contiguity matrix is not transformed into a global normalized spatial weight matrix. Anselin (1995) defined the first local Geary coefficient as

$$C_i^* = \sum_{j=1}^{n} v_{ij}(y_i - y_j)^2 . \tag{21}$$

Suppose that the variable is standardized, and the spatial contiguity matrix is transformed into a row normalized spatial weight matrix. Anselin (1995) defines the second local Geary coefficient as

$$C_i^{**} = \frac{1}{\sigma^2} \sum_{j=1}^{n} w_{ij}^*(y_i - y_j)^2 . \tag{22}$$

Summation of equation (21) divided by the population variance $\sigma^2$ is

$$\frac{1}{\sigma^2} \sum_{i=1}^{n} C_i^* = \frac{n \sum_{i=1}^{n}\sum_{j=1}^{n} v_{ij}(y_i - y_j)^2}{\sum_{i=1}^{n} y_i^2} = \frac{2nV_0}{n-1} \frac{(n-1)\sum_{i=1}^{n}\sum_{j=1}^{n} v_{ij}(y_i - y_j)^2}{2V_0 \sum_{i=1}^{n} y_i^2} = \gamma_c C , \tag{23}$$

where $C$ refers to global Geary coefficient. It can be expressed as

$$C = \frac{(n-1)\sum_{i=1}^{n}\sum_{j=1}^{n} v_{ij}(y_i - y_j)^2}{2V_0 \sum_{i=1}^{n} y_i^2} = \frac{1}{2s^2} \sum_{i=1}^{n}\sum_{j=1}^{n} w_{ij}(y_i - y_j)^2 = \frac{1}{2} \sum_{i=1}^{n}\sum_{j=1}^{n} w_{ij}(z_i^* - z_j^*)^2 . \tag{24}$$

In addition, the proportional coefficient between the sum of the first local Geary coefficient divided by the population variance and the global Geary coefficient is as below

$$\gamma_c = \frac{2nV_0}{n-1} \tag{25}$$

The standardized size variable based on the sample standard deviation s is used here, i.e

$$z_i^* = \frac{y_i}{s} = \frac{x_i - \overline{x}}{s}, \quad z_i^* = \frac{y_j}{s} = \frac{x_j - \overline{x}}{s}.$$

Therefore, the relationship between the sum of the first local Geary coefficients and the global Geary coefficients is

$$\sum_{i=1}^{n} C_i^* = \frac{2nV_0\sigma^2}{n-1} C = \gamma_c \sigma^2 C . \tag{26}$$

This formula is correct, and it satisfies the two requirements given by Anselin (1995). However, it is neither direct nor standard. Dividing the summation of equation (21) by both the population variance $\sigma^2$ and the sum of the spatial weight matrix $V_0$ to obtain the relationship between the local Geary's coefficient and the global Geary coefficient, that is

$$\sum_{i=1}^{n} C_i^{**} = \frac{n \sum_{i=1}^{n} \sum_{j=1}^{n} v_{ij}(y_i - y_j)^2}{V_0 \sum_{i=1}^{n} y_i^2} = \frac{2n}{n-1} \frac{(n-1) \sum_{i=1}^{n} \sum_{j=1}^{n} w_{ij}(y_i - y_j)^2}{2 \sum_{i=1}^{n} y_i^2} = \frac{2n}{n-1} C. \qquad (27)$$

This is different from the relationship between local Geary coefficient and global Geary's coefficient given by Anselin (1995). The reason is that derivation of this relationship is based on the global normalization of spatial weight matrix. Based on the row-normalized weight matrix, the sum of local Geary's coefficients is

$$\sum_{i=1}^{n} C_i^{**} = \frac{n \sum_{i=1}^{n} \frac{1}{V_i} \sum_{j=1}^{n} v_{ij}(y_i - y_j)^2}{\sum_{i=1}^{n} y_i^2} = \frac{n V_0 \sum_{i=1}^{n} \frac{1}{V_i} \sum_{j=1}^{n} w_{ij}(y_i - y_j)^2}{\sum_{i=1}^{n} y_i^2}. \qquad (28)$$

The constant proportional relationship between local Geary coefficient and global Geary coefficient cannot be derived in terms of equation (28). Anselin (1995) believes that, according to equation (25), for the weight matrix normalized by row, $V_0 = n$, so there is $\gamma_c = 2n^2/(n-1)$, that's right. Then he gave the following relation

$$\sum_{i=1}^{n} C_i^{**} = \frac{2n^2}{n-1} C = \gamma_c C. \qquad (29)$$

This is wrong and cannot be strictly derived by mathematical methods, nor can it be verified by observational data. Based on the row-normalized weight matrix, the correct result is

$$\sum_{i=1}^{n} C_i^{**} = \frac{2n}{n-1} \frac{\sum_{i=1}^{n} \sum_{j=1}^{n} w_{ij}^*(z_i^* - z_j^*)^2}{\sum_{i=1}^{n} \sum_{j=1}^{n} w_{ij}(z_i^* - z_j^*)^2} C = \gamma_c^* C, \qquad (30)$$

in which $\gamma_c^*$ represents the proportionality coefficient. The coefficient can be expressed as

$$\gamma_c^* = \frac{2n}{n-1} \frac{\sum_{i=1}^{n} \sum_{j=1}^{n} w_{ij}^*(z_i^* - z_j^*)^2}{\sum_{i=1}^{n} \sum_{j=1}^{n} w_{ij}(z_i^* - z_j^*)^2} \qquad (31)$$

which is not a constant. It cannot be proved that equation (29) is equivalent to equation (30). Moreover, starting from equations (21) and (22), the proportional relationship between the two sets of local Geary coefficients is

$$\frac{C_i^*}{C_i^{**}} = \frac{\sigma^2 \sum_{j=1}^{n} v_{ij}(y_i - y_j)^2}{\sum_{j=1}^{n} \frac{v_{ij}}{V_i}(y_i - y_j)^2} = \sigma^2 V_i = \frac{I_i^*}{I_i^{**}}. \qquad (32)$$

This is obviously not a constant, but a variable that changes with the sum of the rows of the spatial proximity matrix. This shows that the two sets of local Geary coefficients are not equivalent to each other, and the ratio of the corresponding values of the two sets of local Geary coefficients is equal to the ratio of the values of the two sets of local Moran's indices. In short, the second set of local Geary statistic does not satisfy the second requirement given by Anselin (1995).

## 2.2 Revised and normalized results

### 2.2.1 Adjustment of symbol system and clarification of concept

Concept is the cornerstone of logic. If and only the concept is clear, there will be no mistakes in reasoning. The premise of mathematical reasoning is the symbolization of concepts. Confusion of symbols can easily lead to mistakes in reasoning. The main reason for the inconsistency between the two sets of LISA proposed by Anselin (1995) is the unintentional concept substitution caused by the symbol mixing of spatial measure matrixes. At present, there are several problems about spatial autocorrelation in geographical literature.

Firstly, the symbols of spatial contiguity matrix (SCM) and spatial weight matrix (SWM) are confused with each other. The two matrixes are regarded as equivalence and are both represented by the same symbol [$w_{ij}$]. In fact, the spatial distance matrix can be transformed into a spatial contiguity matrix according to a certain distance decay function, and the weight matrix can be obtained by normalizing the spatial contiguity matrix (Chen, 2013; Chen, 2015). Despite the final result is the same in the case of symbol confusion, the form causes many unnecessary misunderstandings for beginners. This paper distinguishes the symbols as follows: SCM is represented by **V**, its elements are represented by $v_{ij}$; SWM is represented by **W**, and its elements are expressed as $w_{ij}$. Thus we have SCM, **V**=[$v_{ij}$], and SWM, **W** =[$w_{ij}$].

Secondly, after the spatial contiguity matrix (SCM) is transformed into the spatial weight matrix (SWM), the global normalization and local normalization by row are confused. Anselin (1995), the original founder of the local Moran index, adopted the method of row normalization (he term the processing "row-standardization"). The sum of the SWM elements is thus equal to *n*. However, this method will lead to two results: (1) The symmetry of the spatial distance matrix is broken. Spatial weight matrix comes from spatial distance matrix or generalized spatial distance matrix. One of the important properties of distance measure is symmetry: $d_{ij}=d_{ji}$ holds for all *i* and *j* (Chen, 2016). This is one of the four principles of the distance axioms (positivity, specification, symmetry, and triangle inequality). (2) The absolute value of the calculated local Moran index may exceed 1 sometimes. Moran index is an autocorrelation coefficient whose absolute value should fall between - 1 and 1 in theory.

Thirdly, the population variance is confused with the sample variance. Moran's index is defined based on population variance, and Geary's coefficient is defined based on sample variance (Chen, 2013). The population variance is expressed as $\sigma^2$, and the denominator in the formula is *n*; the sample variance is expressed as $s^2$, and the denominator in the formula is *n* -1 in the formula. The relationship between them is $\sigma^2=(n-1)s^2/n$.

Fourth, confusion between row summation and column summation. The sum based on row vector is expressed as summation by *j*, and the sum of column vector is expressed as summation by *i*. Based on global normalized weight matrix, the difference is only formal and has nothing to do with the results. However, based on row-normalized weight matrix, the results of row summation differs from the results of column summation.

Fifth, the concepts of normalization and standardization are confused. Generalized standardization includes normalization. However, both standardization and normalization have different definition methods and corresponding calculation formulas. The conversion formula of variables should be determined according to different research objectives.

Table 2 Comparison between Anselin's symbol system and the symbol system in this paper

| Measure set | Anselin | This paper |
| --- | --- | --- |
| Spatial proximity matrix (SPM) | -- | $\mathbf{U}=\{d_{ij}\}$ |
| Spatial contiguity matrix (SCM) | $W=\{w_{ij}\}$ | $\mathbf{V}=\{v_{ij}\}$ |
| Spatial weight matrix (SWM) | $W=\{w_{ij}\}$ | $\mathbf{W}=\{w_{ij}\}$ |
| Sum of elements of spatial contiguity matrix | $S_0$ | $V_0$ |
| Sum of elements of spatial weight matrix | $S_0$ | $W_0$ |
| Size variable | -- | $x_i, x_j$ |
| Centralized variable | $z_i, z_j$ | $y_i, y_j$ |

| Standardized variable | -- | $z_i, z_j$ |
|---|---|---|
| Population variance | $m_2$ | $\sigma^2$ |
| Sample variance | -- | $s^2$ |
| Global Moran's $I$ | $I$ | $I$ |
| Local Moran's $I$ | $I_i$ | $I_i$ |
| Global Geary's $I$ | $c$ | $C$ |
| Local Geary's $I$ | $c_i$ | $C_i$ |

In order to make it easy for readers to understand, I first distinguish symbols, and then clarify the concept of variable transformation. There are three principles for adopting symbols in this paper: First, the principle of consensus. Priority will be given to the conventional expression in the field of mathematics. For example, the population standard deviation is expressed as $\sigma$, and the sample standard deviation is expressed as $s$. Second, the principle of direction. For example, the spatial weight matrix represents **W** because "W" it is the capital form of the initial of "weight". Third, the principle of distinction. For example, the spatial contiguity matrix represents **V**, so as to distinguish it from the spatial weight matrix **W**, and this distinguishing facilitates mathematical reasoning. Among the above three principles, the distinction principle is the most important (Table 2). In the spatial autocorrelation literature, centralization variables (such as defining local Moran's index), standardized variables (such as simplifying the calculation of global Moran index) and global normalized variables (such as simplifying the calculation of Getis-Ord's index) are used, respectively (Table 3). In the literature, when the spatial weight matrix is normalized by row, the concept of row standardization is adopted, but the calculation formula is not given (Anselin, 1995). This can easily lead to misunderstandings for beginners of spatial autocorrelation analysis.

**Table 3 Variable conversion methods, calculation formulas, and properties of converted variables**

| Method | Calculation formula | Property |
|---|---|---|
| Centralization | $y_i = x_i - \bar{x}$ | The mean value is 0 |
| Standardization by z-score | $z_i = (x_i - \bar{x})/\sigma$, $z_i^* = (x_i - \bar{x})/s$, | The mean value is 0 and the standard deviation is 1 |
| Range normalization | $x_i^{(r)} = (x_i - x_{min})/(x_{max} - x_{min})$ | The values range from 0 to 1 |
| Global normalization | $x_i^{(t)} = x_i / \sum_i x_i$, $w_{ij} = v_{ij} / \sum_i \sum_j v_{ij}$ | The values come between 0 and 1 and the sum of the values equals 1 |

### 2.2.2 Definition of normalized local Moran's index

Moran's index is defined on the basis of population standard deviation rather than sample standard deviation. Accordingly, local Moran's index should also be defined through population standard deviation. In light of equation (7), canonical local Moran's index can be defined as

$$I_i = \frac{I_i^*}{\sigma^2 V_0} = \frac{1}{\sigma^2} y_i \sum_{j=1}^{n} \frac{v_{ij}}{V_0} y_j = z_i \sum_{j=1}^{n} w_{ij} z_j. \tag{33}$$

Further, according to equation (7), the relation between global Moran's index and the sum of local Moran's indexes is

$$I = \sum_{i=1}^{n} (\frac{I_i^*}{\sigma^2 V_0}) = \sum_{i=1}^{n} I_i. \tag{34}$$

According to equation (33), the relation between Anselin's first set of local Moran indexes and the local Moran's indexes formula improved in this paper is

$$I_i^* = \gamma I_i = \sigma^2 V_0 I_i. \tag{35}$$

Thus, for the global normalized spatial weight matrix **W** and the standardized variable based on population standard deviation **z**, we have $\sigma^2=1$, $V_0=1$. Thus, equation (9) should be replaced by

$$\gamma_0 = \sigma^2 V_0 = (\frac{1}{n} \sum_{i=1}^{n} z_i^2)(\sum_{i=1}^{n} \sum_{j=1}^{n} w_{ij}) = 1. \tag{36}$$

This suggests that, according to the idea from Anselin (1995), the sum of normalized local Moran's index equals the global Moran's index.

### 2.2.3 Definition of normalized local Geary's coefficient

Geary's coefficient is defined on the basis of sample standard deviation rather than population standard deviation. Accordingly, local Geary's coefficient should also be defined through sample standard deviation. In terms of equation (26), global Geary's coefficient can be expressed as

$$C = \frac{n-1}{2nV_0\sigma^2} \sum_{i=1}^{n} C_i^* = \frac{1}{2V_0 s^2} \sum_{i=1}^{n} C_i^* = \sum_{i=1}^{n} (\frac{C_i^*}{2V_0 s^2}) = \sum_{i=1}^{n} C_i, \tag{37}$$

where $s^2 = n\sigma^2/(n-1)$ reflects the relationship between sample variance $s^2$ and population variance $\sigma^2$. Thus local Geary's coefficient can be defined as

$$C_i = \frac{C_i^*}{2V_0 s^2} = \frac{1}{2V_0 s^2} \sum_{j=1}^{n} v_{ij} (y_i - y_j)^2 = \frac{1}{2} \sum_{j=1}^{n} w_{ij} (z_i^* - z_j^*)^2. \tag{38}$$

Summing equation (38) yields global Geary's coefficient, that is, equation (24). According to equation (37), the relation between Anselin's first set of Geary's coefficient and the local Geary's coefficient formula improved in this paper is

$$C_i^* = \gamma_c \sigma^2 C_i = 2s^2 V_0 C_i. \tag{39}$$

Thus, for the global normalized spatial weight matrix $\mathbf{W}$ and the standardized variable based on sample standard deviation $\mathbf{z}^*$, we have $s^2=1$, $V_0=1$. Thus, according to equation (26), the relation between proportionality coefficients is

$$\gamma_c \sigma^2 = 2s^2 V_0 = 2. \tag{40}$$

Moran's index and Geary's coefficient reflect the same problem from different angles of view. It can be proved that the relationship between global Moran's $I$ and global Geary's $C$ is as follows

$$C = \frac{\sum_{i=1}^{n}\sum_{j=1}^{n} v_{ij} y_i^2 - \sum_{i=1}^{n}\sum_{j=1}^{n} v_{ij} y_i y_j}{V_0 \frac{1}{n-1}\sum_{i=1}^{n} y_i^2} = \frac{n-1}{n}(\mathbf{o}^T \mathbf{W} \mathbf{z}^2 - \mathbf{z}^T \mathbf{W} \mathbf{z}) = \frac{n-1}{n}(\mathbf{o}^T \mathbf{W} \mathbf{z}^2 - I), \tag{41}$$

where $\mathbf{o}^T = [1\ 1\ \ldots\ 1]$ is a row vector in which the elements are all 1. The symbol "T" indicates transposition. If the mean of the global Moran's index is treated as $I_0 = 1/(1-n)$, the mean of global Geary's coefficient, $C_0$, can be estimated by

$$C_0 = \frac{n-1}{n}(\mathbf{e}^T \mathbf{W} \mathbf{z}^2 - I_0) = \frac{n-1}{n}(\mathbf{e}^T \mathbf{W} \mathbf{z}^2 - \frac{1}{1-n}) = \frac{n-1}{n}\mathbf{e}^T \mathbf{W} \mathbf{z}^2 + \frac{1}{n}. \tag{42}$$

Further, the relationship between local Moran's indexes and local Geary's coefficient can be derived. From equation (38) it follows

$$C_i = \frac{1}{2V_0 \frac{n}{n-1}\sigma^2} \sum_{j=1}^{n} v_{ij}(y_i - y_j)^2 = \frac{n-1}{2n}\sum_{j=1}^{n} w_{ij}(z_i - z_j)^2. \tag{43}$$

Changing the form of equation (43) yields

$$C_i = \frac{n-1}{2n}(\sum_{j=1}^{n} w_{ij}(z_i^2 + z_j^2) - 2\sum_{j=1}^{n} w_{ij} z_i z_j) = \frac{n-1}{2n}(\sum_{j=1}^{n} w_{ij}(z_i^2 + z_j^2) - 2I_i). \tag{44}$$

This means that there is a strict numerical conversion relationship between local Moran's indexes and local Geary's coefficient, although they describe the same problem from different angles. It can be seen that equation (41) can be obtained by summing equation (44).

# 3 Empirical analysis

## 3.1 Study area and data

Taking cities in Beijing, Tianjin and Hebei (BTH) region as an example, a concise calculation case is given in this section. This is a demonstrative case, not an explanatory case. In other words, this example is used to verify the reasoning results rather than to study the spatial structure and characteristics of BTH urban systems. The study area includes Beijing city, Tianjin city, and the main cities of Hebei Province (Figure 1). The study region is also termed Jing-Jin-Ji (JJJ) region in literature. The cities are all of prefecture level and above, and the number of cities is $n = 13$. The size measurement is the city population of the fifth census in 2000 and the sixth census in 2010. Town population is not taken into account. At present, urban population has the definitions of regional total population, municipal population, city population and urban population consisting city population and town population. This case uses the city population, which can better reflect the characteristics of city size. The population size was processed by centralization ($y$), population-based standardization ($z$) and sample-based standardization ($z^*$) (Table 4). As for the spatial weight matrix, the basic data is derived from the traffic mileage between cities (Table 5). The spatial weight function adopts the special negative power law, the inverse proportion function, which is actually the intersection of power law and hyperbolic function. Thus, the spatial contiguity is defined as

$$v_{ij} = \begin{cases} 1/d_{ij}, & i \neq j \\ 0, & i = j \end{cases}, \tag{45}$$

where $d_{ij}$ denotes the distance by road between city $i$ and city $j$. On this basis, the traffic mileage matrix (**U**) can be transformed into a spatial contiguity matrix (**V**), which can be changed to the global normalization weight matrix (**W**) and row normalization weight matrix (**W**$^*$).

**Table 4 Beijing-Tianjin-Hebei city population and its centralization and standardization results**

| City | 2000 | | | | 2010 | | | |
|---|---|---|---|---|---|---|---|---|
| | $x$ | $y$ | $z$ | $z^*$ | $x$ | $y$ | $z$ | $z^*$ |
| **Beijing** | 949.6688 | 769.1377 | 2.9976 | 2.8800 | 1555.2378 | 1284.2528 | 2.9870 | 2.8698 |
| **Tianjin** | 531.3702 | 350.8391 | 1.3673 | 1.3137 | 885.6234 | 614.6384 | 1.4296 | 1.3735 |
| **Shijiazhuang** | 193.0579 | 12.5268 | 0.0488 | 0.0469 | 275.6871 | 4.7021 | 0.0109 | 0.0105 |
| **Tanshan** | 140.3887 | -40.1424 | -0.1564 | -0.1503 | 163.7579 | -107.2271 | -0.2494 | -0.2396 |
| **Qinhuangdao** | 70.7267 | -109.8044 | -0.4279 | -0.4112 | 95.1872 | -175.7978 | -0.4089 | -0.3928 |
| **Handan** | 107.1068 | -73.4243 | -0.2862 | -0.2749 | 111.7417 | -159.2433 | -0.3704 | -0.3558 |
| **Xingtai** | 53.6282 | -126.9029 | -0.4946 | -0.4752 | 63.7797 | -207.2053 | -0.4819 | -0.4630 |
| **Baoding** | 90.2496 | -90.2815 | -0.3519 | -0.3381 | 98.0177 | -172.9673 | -0.4023 | -0.3865 |
| **Zhangjiakou** | 79.6580 | -100.8731 | -0.3931 | -0.3777 | 90.0218 | -180.9632 | -0.4209 | -0.4044 |
| **Chengde** | 32.5821 | -147.9490 | -0.5766 | -0.5540 | 49.8293 | -221.1557 | -0.5144 | -0.4942 |
| **Cangzhou** | 44.3561 | -136.1750 | -0.5307 | -0.5099 | 48.9701 | -222.0149 | -0.5164 | -0.4961 |
| **Langfang** | 29.5879 | -150.9432 | -0.5883 | -0.5652 | 46.6539 | -224.3311 | -0.5218 | -0.5013 |
| **Hengshui** | 24.5229 | -156.0082 | -0.6080 | -0.5842 | 38.2976 | -232.6874 | -0.5412 | -0.5200 |
| **Mean** | **180.5311** | **0.0000** | **0.0000** | **0.0000** | **270.9850** | **0.0000** | **0.0000** | **0.0000** |
| **σ** | **256.5845** | **256.5845** | **1.0000** | **0.9608** | **429.9496** | **429.9496** | **1.0000** | **0.9608** |
| **s** | **267.0616** | **267.0616** | **1.0408** | **1.0000** | **447.5057** | **447.5057** | **1.0408** | **1.0000** |

**Table 5 Spatial distance matrix of Beijing-Tianjin-Hebei cities based on traffic mileage**

| City | Beijing | Tianjin | Shijiazhuang | Tanshan | Qinhuangdao | Handan | Xingtai | Baoding | Zhangjiakou | Chengde | Cangzhou | Langfang | Hengshui |
|---|---|---|---|---|---|---|---|---|---|---|---|---|
| Beijing | 0 | 160.8855 | 321.7625 | 185.4770 | 288.9055 | 479.9810 | 430.2520 | 187.1300 | 198.1975 | 194.5940 | 233.4440 | 83.2755 | 299.7580 |
| Tianjin | 160.8855 | 0 | 344.5825 | 101.4105 | 242.6355 | 454.8400 | 425.3890 | 201.9420 | 332.9375 | 280.6470 | 138.6135 | 86.1555 | 259.8555 |
| Shijiazhuang | 321.7625 | 344.5825 | 0 | 423.7510 | 568.1560 | 167.2815 | 114.0840 | 138.9090 | 430.8215 | 506.6400 | 221.7565 | 283.2495 | 142.5935 |
| Tanshan | 185.4770 | 101.4105 | 423.7510 | 0 | 151.3880 | 547.4205 | 517.8910 | 289.5120 | 376.8000 | 185.3500 | 215.0285 | 144.6130 | 352.4360 |
| Qinhuangdao | 288.9055 | 242.6355 | 568.1560 | 151.3880 | 0 | 711.7120 | 662.2960 | 433.9170 | 481.3360 | 222.2030 | 375.5205 | 292.9180 | 508.4835 |
| Handan | 479.9810 | 454.8400 | 167.2815 | 547.4205 | 711.7120 | 0 | 53.4600 | 296.7465 | 606.6940 | 664.8585 | 335.0465 | 440.4685 | 214.2995 |
| Xingtai | 430.2520 | 425.3890 | 114.0840 | 517.8910 | 662.2960 | 53.4600 | 0 | 245.8830 | 557.3515 | 615.1295 | 299.4430 | 391.1260 | 167.0325 |
| Baoding | 187.1300 | 201.9420 | 138.9090 | 289.5120 | 433.9170 | 296.7465 | 245.8830 | 0 | 278.0950 | 372.0075 | 150.5130 | 147.8300 | 144.8405 |
| Zhangjiakou | 198.1975 | 332.9375 | 430.8215 | 376.8000 | 481.3360 | 606.6940 | 557.3515 | 278.0950 | 0 | 372.8730 | 411.7425 | 257.5700 | 455.2955 |
| Chengde | 194.5940 | 280.6470 | 506.6400 | 185.3500 | 222.2030 | 664.8585 | 615.1295 | 372.0075 | 372.8730 | 0 | 407.1040 | 259.8085 | 495.3555 |
| Cangzhou | 233.4440 | 138.6135 | 221.7565 | 215.0285 | 375.5205 | 335.0465 | 299.4430 | 150.5130 | 411.7425 | 407.1040 | 0 | 149.7245 | 140.0620 |
| Langfang | 83.2755 | 86.1555 | 283.2495 | 144.6130 | 292.9180 | 440.4685 | 391.1260 | 147.8300 | 257.5700 | 259.8085 | 149.7245 | 0 | 237.8790 |
| Hengshui | 299.7580 | 259.8555 | 142.5935 | 352.4360 | 508.4835 | 214.2995 | 167.0325 | 144.8405 | 455.2955 | 495.3555 | 140.0620 | 237.8790 | 0 |

## 3.2 Calculation results

For the data of two years and two statistics, i.e., local Moran index and local Geary coefficient, three sets of calculation results are given, respectively. For the local spatial statistics defined by Anselin (1995), the first set of local Moran index is expressed as MI1, the second set of local Moran index as MI2; the first set of local Geary coefficients is expressed as GC1, and the second set of local Geary coefficients is written as GC2. Accordingly, the modified local Moran index and Geary coefficient are expressed as MI3 and GC3, respectively. The results are as follows. First, the ratio of MI1 to MI2 is not a constant, and the ratio of GC1 to GC2 is neither a constant. This proves that the two sets of local Moran indices and the two sets of local Geary coefficients of Anselin (1995) are not equivalent to one another; Secondly, the ratio of MI1 to MI3 is a constant, and the ratio of GC1 to GC3 is also a constant. It is proved that the first set of local Moran index of Anselin (1995) is equivalent to the modified local Moran index in this paper, and the first set of local Geary coefficient of Anselin (1995) is also equivalent to the modified local Geary coefficient of this paper (Table 6, Table 7). The reason is that the first set of local Moran index and local Geary coefficient defined by Anselin (1995) are based on symmetric spatial contiguity matrix. The modified statistics in this paper are based on the global normalized spatial weight matrix which is symmetric, while the second set of local Moran index and local Geary coefficient defined by Anselin (1995) are based on the local normalized spatial weight matrix, in which the symmetry is broken.

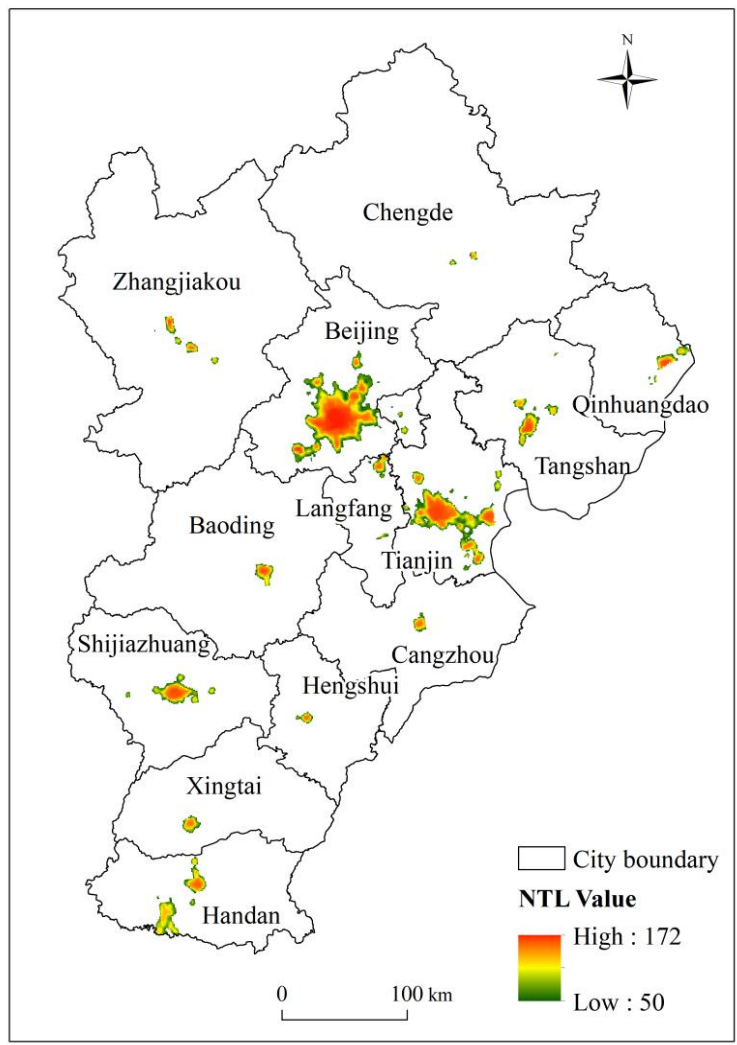

**Figure 1 Main cities in Beijing, Tianjin, and Hebei region, China**

Using the calculation results, we can verify two key equations. The relationship between the sum of the first set of local Moran indexes and the global Moran index satisfies equation (8), and the relationship between the sum of the first set of local Geary coefficients and the global Geary coefficient satisfies equation (26). However, the relationship between the sum of the second set of local Moran indexes and the global Moran index does no satisfy equation (16), and the relationship between the sum of the second set of local Geary coefficients and the global Geary coefficient does not satisfy equation (27). The sum of spatial contiguity matrices is $V_0$=0.6671. In 2000, the population variance of city population in Beijing-Tianjin-Hebei region is $\sigma^2$=65835.5974, thus $\gamma=\sigma^2 V_0$=43916.8725, the global Moran index is $I$=-0.1191, and the sum of the first set of local Moran indexes is $\sum I_i^*$=-5229.3702=$\gamma I$=43916.8725*(-0.1191). On the other hand, $n$=13, $\gamma_c=2nV_0/(n-1)$=1.4453, and the global Geary coefficient is $C$=1.1377, so the sum of the first set of local Geary coefficients is $\sum C_i^*$=108253.8824=$\gamma_c\sigma^2 C$=1.4453*65835.5974*1.1377. However, the sum of the second set of local Moran indices is $\sum I_i^{**}$=-1.4299, while $n*I$=13*(-0.1191)=-1.5480. The two values are not equal to one another (-1.4299≠-1.5480). The sum of the second set of local Geary coefficients is $\sum C_i^{**}$=30.4883, and $2n^2*C/(n-1)$=28.1667*1.1377=32.0446. The two values are not equal to one another (30.4883≠32.0446). The sum of the third set of local Moran index is equal to the global Moran index, the ratio of the first set of local Moran indexes to the corresponding third set of local Moran indexes is $\gamma=\sigma^2 V_0$=43916.8725, which is a constant; the sum of the third set of local Geary coefficients equals the global Geary coefficient, and the ratio of the first set of local Geary coefficients to the corresponding third set of local Geary coefficient is $\gamma_c\sigma^2$= 1.4453* 65835.5974 = 95153.2237 is a constant (Table 6, Table 7).

If the calculation result of one year is an isolated case, we might as well take a look at the situation in 2010. Based on the 6[th] census data, the population variance of Beijing-Tianjin-Hebei city population is $\sigma^2$=184856.6464, thus $\gamma=\sigma^2 V_0$=123312.1000, the global Moran index is $I$=-0.1124, and the sum of the first set of local Moran indexes is $\sum I_i^*$=-13856.5039=$\gamma I$=123312.1000*(-0.1124). On the other hand, $\gamma_c$= 1.4453, and the global Geary coefficient is $C$=1.1329, so the sum of the first set of local Geary coefficients is $\sum C_i^*$=302682.5671 = $\gamma_c\sigma^2 C$ = 1.4453*184856.6464*1.1329. However, the sum of the second set of local Moran indices is $\sum I_i^{**}$=-1.3523, while $n*I$=13*(-0.1124)=-1.4608 (Figure 2(a)). The two numbers are not equal to each other (-1.3523≠-1.4608). The sum of the second set of local Geary coefficients is $\sum C_i^{**}$=30.3506, and $2n^2*C/(n-1)$ = 28.1667*1.1329 =31.9099. The two numbers are not equal to each other (30.3506≠31.9099). The sum of the third

set of local Moran index is equal to the global Moran index, the ratio of the first set of local Moran indexes to the corresponding numbers in the third set of local Moran index is $\gamma = \sigma^2 V_0 = 123312.1000$ (Figure 2(b)); the sum of the third set of local Geary coefficients equals the global Geary coefficient, and the ratio of the first set of local Geary coefficient to the corresponding third set of local Geary coefficient is $\gamma_c \sigma^2 = 1.4453 * 184856.6464 = 267176.2168$ is a constant (Table 6, Table 7). It can be seen that the calculation results of the two years fully support the previous theoretical conclusions and related judgments.

Table 6 Comparison of three sets of local Moran index values in two years

| City | 2000 | | | | | 2010 | | | | |
|---|---|---|---|---|---|---|---|---|---|---|
| | Local MI1 | Local MI2 | Local MI3 | MI1/MI2 | MI1/MI3 | Local MI1 | Local MI2 | Local MI3 | MI1/MI2 | MI1/MI3 |
| Beijing | -2686.4966 | -0.7067 | -0.0612 | 3801.3644 | 43916.8725 | -7140.4536 | -0.6690 | -0.0579 | 10673.67042 | 123312.1000 |
| Tianjin | -387.0133 | -0.0951 | -0.0088 | 4071.1117 | 43916.8725 | -1175.2192 | -0.1028 | -0.0095 | 11431.08104 | 123312.1000 |
| Shijiazhuang | -23.1481 | -0.0068 | -0.0005 | 3385.2705 | 43916.8725 | -14.4935 | -0.0015 | -0.0001 | 9505.340198 | 123312.1000 |
| Tanshan | -121.7919 | -0.0343 | -0.0028 | 3547.3310 | 43916.8725 | -603.5770 | -0.0606 | -0.0049 | 9960.382257 | 123312.1000 |
| Qinhuangdao | -142.9763 | -0.0607 | -0.0033 | 2356.2158 | 43916.8725 | -379.2385 | -0.0573 | -0.0031 | 6615.906335 | 123312.1000 |
| Handan | 170.5561 | 0.0533 | 0.0039 | 3202.3026 | 43916.8725 | 594.8129 | 0.0662 | 0.0048 | 8991.593275 | 123312.1000 |
| Xingtai | 185.0124 | 0.0511 | 0.0042 | 3618.1153 | 43916.8725 | 637.3519 | 0.0627 | 0.0052 | 10159.13409 | 123312.1000 |
| Baoding | -92.0058 | -0.0244 | -0.0021 | 3771.5181 | 43916.8725 | -335.7750 | -0.0317 | -0.0027 | 10589.86662 | 123312.1000 |
| Zhangjiakou | -231.9379 | -0.1057 | -0.0053 | 2194.2630 | 43916.8725 | -708.7104 | -0.1150 | -0.0057 | 6161.166944 | 123312.1000 |
| Chengde | -363.3994 | -0.1476 | -0.0083 | 2461.9446 | 43916.8725 | -889.9662 | -0.1287 | -0.0072 | 6912.777246 | 123312.1000 |
| Cangzhou | -194.7349 | -0.0538 | -0.0044 | 3620.4838 | 43916.8725 | -561.9455 | -0.0553 | -0.0046 | 10165.78443 | 123312.1000 |
| Langfang | -1369.3138 | -0.3073 | -0.0312 | 4455.7783 | 43916.8725 | -3399.6518 | -0.2717 | -0.0276 | 12511.16811 | 123312.1000 |
| Hengshui | 27.8793 | 0.0081 | 0.0006 | 3431.1735 | 43916.8725 | 120.3620 | 0.0125 | 0.0010 | 9634.229089 | 123312.1000 |
| Sum | -5229.3702 | -1.4299 | -0.1191 | 43916.8725 | 570919.3421 | -13856.5039 | -1.3523 | -0.1124 | 123312.1000 | 1603057.3005 |
| Expected | -5229.3702 | -1.5480 | -0.1191 | 43916.8725 | 570919.3421 | -13856.5039 | -1.4608 | -0.1124 | 123312.1000 | 1603057.3005 |

Table 7 Comparison of three sets of local Geary coefficient values in two years

| City | 2000 | | | | | 2010 | | | | |
|---|---|---|---|---|---|---|---|---|---|---|
| | Local GC1 | Local GC2 | Local GC3 | GC1/GC2 | GC1/GC3 | Local GC1 | Local GC2 | Local GC3 | GC1/GC2 | GC1/GC3 |
| Beijing | 41036.8054 | 10.7953 | 0.4313 | 3801.3644 | 95153.2237 | 113754.5272 | 10.6575 | 0.4258 | 10673.6704 | 267176.2168 |
| Tianjin | 12819.0307 | 3.1488 | 0.1347 | 4071.1117 | 95153.2237 | 37929.2182 | 3.3181 | 0.1420 | 11431.0810 | 267176.2168 |
| Shijiazhuang | 2908.7705 | 0.8592 | 0.0306 | 3385.2705 | 95153.2237 | 8029.3420 | 0.8447 | 0.0301 | 9505.3402 | 267176.2168 |
| Tanshan | 5340.6947 | 1.5056 | 0.0561 | 3547.3310 | 95153.2237 | 15962.5572 | 1.6026 | 0.0597 | 9960.3823 | 267176.2168 |
| Qinhuangdao | 3628.6681 | 1.5400 | 0.0381 | 2356.2158 | 95153.2237 | 10073.4191 | 1.5226 | 0.0377 | 6615.9063 | 267176.2168 |
| Handan | 2044.0978 | 0.6383 | 0.0215 | 3202.3026 | 95153.2237 | 5920.6445 | 0.6585 | 0.0222 | 8991.5933 | 267176.2168 |
| Xingtai | 2655.7337 | 0.7340 | 0.0279 | 3618.1153 | 95153.2237 | 7227.0101 | 0.7114 | 0.0270 | 10159.1341 | 267176.2168 |
| Baoding | 5080.6946 | 1.3471 | 0.0534 | 3771.5181 | 95153.2237 | 14731.9805 | 1.3911 | 0.0551 | 10589.8666 | 267176.2168 |
| Zhangjiakou | 4499.9163 | 2.0508 | 0.0473 | 2194.2630 | 95153.2237 | 12851.4607 | 2.0859 | 0.0481 | 6161.1669 | 267176.2168 |

| | | | | | | | | | |
|---|---|---|---|---|---|---|---|---|---|
| **Chengde** | 5353.0964 | 2.1743 | 0.0563 | 2461.9446 | 95153.2237 | 14332.0819 | 2.0733 | 0.0536 | 6912.7772 | 267176.2168 |
| **Cangzhou** | 5400.0965 | 1.4915 | 0.0568 | 3620.4838 | 95153.2237 | 15101.1057 | 1.4855 | 0.0565 | 10165.7844 | 267176.2168 |
| **Langfang** | 13324.4547 | 2.9904 | 0.1400 | 4455.7783 | 95153.2237 | 35822.5797 | 2.8632 | 0.1341 | 12511.1681 | 267176.2168 |
| **Hengshui** | 4161.8231 | 1.2129 | 0.0437 | 3431.1735 | 95153.2237 | 10946.6401 | 1.1362 | 0.0410 | 9634.2291 | 267176.2168 |
| **Sum** | 108253.8824 | 30.4883 | 1.1377 | 43916.8725 | 1236991.9079 | 302682.5671 | 30.3506 | 1.1329 | 123312.1000 | 3473290.8178 |
| **Expected** | 108253.8824 | 32.0446 | 1.1377 | 43916.8725 | 1236991.9079 | 302682.5671 | 31.9099 | 1.1329 | 123312.1000 | 3473290.8178 |

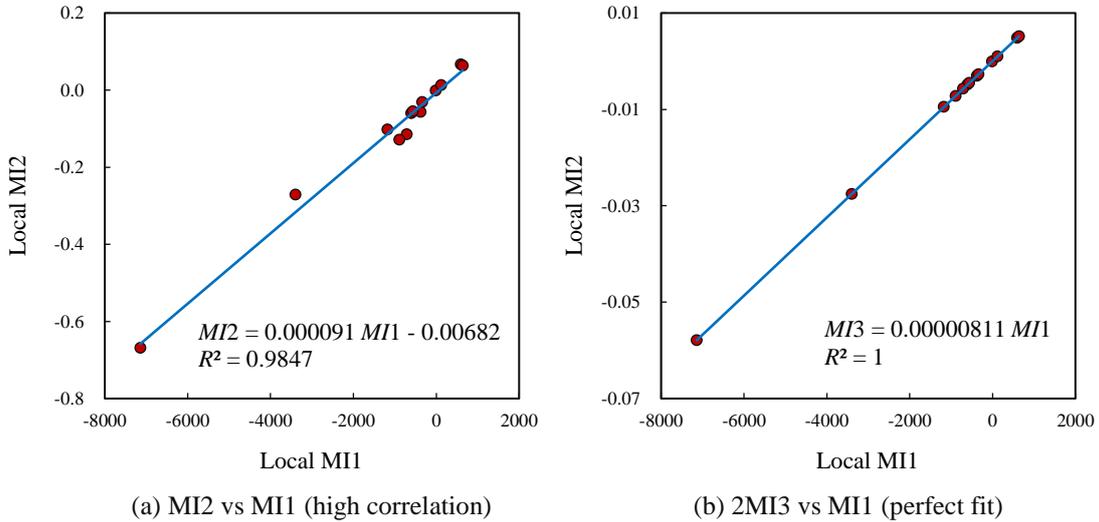

(a) MI2 vs MI1 (high correlation)    (b) 2MI3 vs MI1 (perfect fit)

**Figure 2 The relationships between three sets of local Moran's indexes of BTH cities in 2010**

(**Note**: The second set of local Moran's indexes (MI2) are highly correlated with the first local Moran's indexes (MI1), but not equivalent to one another. The third set of local Moran's indexes (MI3) is equivalent to the first set of local Moran's indexes (MI1). The coefficient $1/\gamma$ = 1/123312.1000=0.000008110. MI2 does not satisfy the second requirement for LISAs given by Anselin (1995).)

# 4 Questions and discussion

The local Moran indexes and the local Geary coefficients given in this paper are derived from Anselin's correct definition and relationship, without substantial innovation. The contribution of this paper lies in two aspects. First, it clarifies a series of logical misunderstandings of local spatial autocorrelation statistics and gives the correct expressions. Second, it normalizes the local spatial autocorrelation statistics, and the canonical results are helpful for more convenient application. If the spatial contiguity matrix is normalized by row, the spatial weight matrix will be asymmetric. Substitution of symmetric spatial weight matrix with asymmetric spatial weight matrix leads to two wrong relations: First, the sum of local Moran index based on standardized variable and local normalized matrix is equal to $n$ times of global Moran index; Second, the sum of local Geary coefficients based on standardized variable and local normalized weight matrix is equal to $2n^2/(n-1)$ times of global Geary coefficient. In fact, the two relations can never be derived from Anselin's hypotheses.

The errors based on the wrong relations are not too significant in many cases, but the results have a far-reaching impact on geographical analysis. Concretely speaking, these incorrect relationships lead to a series of problems (Table 8): (1) The relationship between the definitions of two local Moran indexes is broken (not equivalent to each other). The first set of local LISA is based on symmetric spatial adjacency matrix, and the second set is based on asymmetric spatial weight matrix normalized by row. As a result, the ratio of the values of the two sets of parameters is not a constant. (2) When defining the local spatial autocorrelation index, we only consider the relationship between one element and other elements. However, the pairwise correlation between all elements is ignored. For the local index of the $i$th geographical element, only the relationships between element $i$ and element $j$ are taken into account, the relationships between element $j$ and element $k$ are neglected ($i$, $j$, $k=1,2,3,…,n$). In this case, the wholeness of a geographical system is overlooked in the local spatial analysis. (3) The absolute value of the local Moran index may exceed 1, thus decoupling from the concept of correlation coefficient. Moran's index was proposed by analogy with Pearson correlation. The values of Moran's index comes between -1 and 1. (4) The parameters are lack of clear boundary value and critical value. The boundary values of Moran index is -1 and 1. The critical value is 0 in theory and $1/(1-n)$ in experience. The boundary values of the Geary coefficient are 0 and 2, and the critical value is theoretically 1. In addition, Anselin (1995) used the population standard deviation to replace the sample standard deviation when defining the local Geary coefficient. Where logic is concerned, no problem; while where history is concerned, there is problem: the result violates the original intention of the definition of Geary coefficient. Moran's index, which is derived from Pearson correlation coefficient, as indicated above, is a statistics based on population standard deviation. Geary's coefficient is defined by analogy with Durbin-Watson statistics based on sample standard deviation in order to make up for the deficiency of Moran's index. To define the local Geary coefficient, we should respect the original meaning of the definition of the Geary coefficient, so that the local Geary coefficient can be effectively associated with the global Geary coefficient. From the existing literature, some readers have found Anselin's mistakes. Some scholars adopt a compromise approach. For example, they use the global normalized spatial weight matrix instead of the local normalized spatial weight matrix by row, but multiply $n$ in front of the corrected local Moran index calculation formula[1]. This ensures that the sum of local Moran indexes is equal to *n* times the global Moran index.

Table 8 Functions and problems of Anselin's LISA and the improved effect of this paper

| Definer | Variable | Statistic | Function | Advantages and disadvantages |
|---|---|---|---|---|
| **Anselin** | Central variable and non-normalized symmetric contiguity matrix | First local Moran's $I$ | Reflect local spatial dependence | Simple but lack of clear boundary value and critical value (reference value) |
| | | First local Geary's $C$ | Reflect local spatial dependence | Simple but lack of clear boundary value and critical value (reference value) |
| | Standard variable and row-normalized asymmetric weight matrix | Second Moran's $I$ | Reflect local spatial dependence from the perspective of population | Decoupled from the first definition of local Moran's $I$; Decoupling from correlation coefficient; The relationships between two elements in the system is ignored |
| | | Second Geary's $C$ | Reflect local spatial dependence from the perspective of population | Decoupled from the first definition of local Geary's $C$; Decoupling from the analogy with the Durbin-Watson statistic; The relationships between two elements in the system is ignored; sample standard deviation is replaced by population standard deviation |
| **This paper** | Standardized variable and global normalized symmetric weight matrix | Third Moran's $I$ | Reflect local spatial dependence from the perspective of population | Equivalent to the first definition of local Moran's $I$; Linked to correlation coefficient; The spatial relationship of other elements other than the target geographical elements is considered; There are clear boundary values and critical values |
| | | Third Geary's $C$ | Reflect local spatial dependence from the perspective of samples | Equivalent to the first definition of local Geary's $C$; Linked to generalized Durbin-Watson statistics; The spatial relationship of other elements other than the target geographical elements is considered; Return to the sample analysis perspective of global Geary coefficient; There are clear boundary values and critical values |

---

[1] I found this kind of treatment in some teaching courseware.

As we know, Anselin is a well-known outstanding scholar is the field of geographical spatial analysis. Due to the far-reaching influence of Anselin's work, its logical errors caused confusion in its application and interpretation. Science respects logic and facts, not authority -- only pseudoscience starts from authoritative judgment. In order to solve the above problems, this paper carries out the following processing in the process of mathematical deduction: First, return to the essence of the spatial distance matrix behind the spatial weight matrix, and respect the basic distance axiom. The global spatial weight matrix is obtained by global normalization of spatial contiguity matrix. The global normalized spatial weight matrix is used to replace Anselin's row-normalized weight matrix. In this way, the connotation of the concept before and after is unified and the logic is consistent, so as to avoid reasoning mistakes. Second, start from the original idea of Moran index and Geary coefficient. The normalized local Moran index is defined, and the population standard deviation is used to standardize the size variable; the normalized local Geary coefficient is defined, and the sample standard deviation is used to standardize the size variable. Third, start from the original intention of Anselin. Anselin (1995) gives two sets of local Moran index and local Geary coefficient. We absolutely don't want the inconsistency between them. By examining the reasoning process, we found that the reason for the error lies in the logic error caused by the unintentional concept replacement. According to the sign system and simplification principle of this paper, we transform Anselin's second set of local Moran index and local Geary coefficient formulae. Comparing the two sets of results, we can see the problems and thus understand the similarities and differences between the two sets of formulae (Table 8, Table 9).

**Table 9 Comparison of between normalized LISA and the equivalent transformation results of Anselin's second set of LISA definitions**

| Category | Measure | Definition in this paper | Anselin's definition |
|---|---|---|---|
| Moran's $I$ | Global Moran's $I$ | $I = \sum_{i=1}^{n}\sum_{j=1}^{n} w_{ij} z_i z_j = \mathbf{z}^{\mathrm{T}} \mathbf{W} \mathbf{z}$ | $I = \sum_{i=1}^{n}\sum_{j=1}^{n} w_{ij} z_i z_j$ |
| | Local Moran's $I$ | $I_i = z_i \sum_{j=1}^{n} w_{ij} z_j$ | $I_i = \dfrac{z_i}{S_i} \sum_{j=1}^{n} v_{ij} z_j$ |
| | Sum of local Moran's I | $\sum_{i=1}^{n} I_i = I$ | $\sum_{i=1}^{n} I_i \approx nI$ |

| Geary's $C$ | Global Geary's $C$ | $C = \dfrac{1}{2}\sum\limits_{i=1}^{n}\sum\limits_{j=1}^{n} w_{ij}(z_i^* - z_j^*)^2$ | $C = \dfrac{1}{2}\sum\limits_{i=1}^{n}\sum\limits_{j=1}^{n} w_{ij}(z_i^* - z_j^*)^2$ |
|---|---|---|---|
| | Local Geary's $C$ | $C_i = \dfrac{1}{2}\sum\limits_{j=1}^{n} w_{ij}(z_i^* - z_j^*)^2$ $= \dfrac{n-1}{2n}(\sum\limits_{j=1}^{n} w_{ij}(z_i^2 + z_j^2) - 2I_i)$ | $C_i = \dfrac{1}{S_i}\sum\limits_{j=1}^{n} v_{ij}(z_i - z_j)^2$ |
| | Sum of Local Geary's $C$ | $\sum\limits_{i=1}^{n} C_i = C = \dfrac{n-1}{n}(\mathbf{e}^\mathrm{T}\mathbf{Wz}^2 - I)$ | $\sum\limits_{i=1}^{n} C_i \approx \dfrac{2n^2}{n-1}C$ |

**Note**: For comparison, Anselin's definitions are transformed and re-expressed with new symbols. However, the new expressions are completely equivalent to Anselin's original expressions.

# 5 Conclusions

The global spatial autocorrelation coefficients reflect the sum of any two geographical elements in a region, while the local spatial autocorrelation indexes reflect the sum of correlation between a geographical element and all other geographical elements. The sum of parts is proportional to the whole. The first set of local Moran indexes and Geary coefficients defined by Anselin (1996) is effective and consistent with the idea of global Moran index and Geary coefficient. However, the second set of local Moran indexes and local Geary coefficients defined by him are not equivalent to the first set of parameters. This paper is devoted to correcting the mistakes in its reasoning process and gives the third set of definitions of local Moran indexes and local Geary coefficient in canonical forms. The new local Moran index and local Geary coefficient are simple and concise. The new expressions are consistent with the original intention of Anselin and the statistical essence of global Moran index and global Geary coefficient.

The main points of this paper are summarized as follows.

**Firstly, the LISA defined by Anselin (1995) is of great significance to the analysis of local spatial autocorrelation, but there are also some faults**. The first set of LISA is based on the definition of centralized variables and non-normalized spatial contiguity matrix, lacking clear boundary values and critical value. The second set of local LISA is based on the definitions of standardized variables and row-normalized spatial weight matrix, which ignores the global relationship behind the local analysis. One of the results is that the two sets of indexes are not equivalent to one another. In addition, the population standard deviation is adopted when defining

the second local Geary coefficients, which violates the original intention of Geary coefficient. All the indexes lack clear boundary values and critical value, and they are uncoupled from the correlation coefficient. One consequence is that the analysis process is complex; the other is that the conclusions drawn from the two sets of indexes are often inconsistent with each other. **Secondly, the LISA expression is reconstructed by using the global normalized spatial weight matrix and standardized size variables based on *z*-score to eliminate the defects of Anselin's LISA definition**. By doing so, we have canonical spatial autocorrelation measurements. The global normalized spatial weight matrix is used to replace the row-based local normalized spatial weight matrix. The population standard deviation is used to standardize the variables when defining the local Moran indexes, and the sample standard deviation is used to standardize the variables when defining the local Geary coefficient. The local LISA problem of Anselin can be solved effectively and the results are more concise and simpler. The results given in this paper are equivalent to those given by Anselin's first set of formulas, i.e. first sets of local Moran index and local Geary coefficient, but they are not linearly proportional to the results of the second set of formulas, namely the second sets of local Moran index and local Geary coefficient.

## Acknowledgement:

This research was sponsored by the National Natural Science Foundation of China (Grant No. 42171192). The support is gratefully acknowledged.

## Supplementary Files

**[Supplementary File S1]** *Spatial data sets and calculation results of local spatial autocorrelation indexes for 2000 (Excel)*. This file includes the dataset of spatial distances and city population in 2000, global Moran's indexes and Geary's coefficients, three sets of local Moran's index, and three sets of local Geary's coefficients. The original data and calculation process are displayed for readers.

**[Supplementary File S2]** *Spatial data sets and calculation results of local spatial autocorrelation indexes for 2010 (Excel)*. This file includes the dataset of spatial distances and city population in 2010, global Moran's indexes and Geary's coefficients, three sets of local Moran's index, and three sets of local Geary's coefficients. All the results are tabulated for comparison and references.

# References

Anselin L (1995). Local indicators of spatial association—LISA. *Geographical Analysis*, 27(2): 93–115

Anselin L (1996). The Moran scatterplot as an ESDA tool to assess local instability in spatial association. In: Fischer M, Scholten HJ, Unwin D (eds.). *Spatial Analytical Perspectives on GIS*. London: Taylor & Francis, pp111-125

Chen YG (2012). On the four types of weight functions for spatial contiguity matrix. *Letters in Spatial and Resource Sciences*, 5(2): 65-72

Chen YG (2013). New approaches for calculating Moran's index of spatial autocorrelation. *PLoS ONE*, 8(7): e68336

Chen YG (2015). A new methodology of spatial cross-correlation analysis. *PLoS ONE*, 10(5): e0126158

Chen YG (2016). Spatial autocorrelation approaches to testing residuals from least squares regression. *PLoS ONE*, 11(1): e0146865

Chen YG (2017). An analytical method of growth quotients for understanding industrial development. *Human Geography*, 32(4): 86-94 [In Chinese]

Cliff AD, Ord JK (1973). *Spatial Autocorrelation*. London: Pion Limited

Cliff AD, Ord JK (1981). *Spatial Processes: Models and Applications*. London: Pion Limited

Fotheringham AS (1997). Trends in quantitative methods I: Stressing the Local. *Progress in Human Geography*, 21: 88-96

Fotheringham AS (1998). Trends in quantitative method II: Stressing the computational. *Progress in Human Geography*, 22: 283-292

Fotheringham AS (1999). Trends in quantitative methods III: Stressing the visual. *Progress in Human Geography*, 23(4): 597-606

Geary RC (1954). The contiguity ratio and statistical mapping. *The Incorporated Statistician*, 1954, 5: 115–145

Getis A (2009). Spatial weights matrices. *Geographical Analysis*, 41 (4): 404–410

Getis A, Aldstadt J (2004). Constructing the spatial weights matrix using a local statistic. *Geographical Analysis*, 36 (2): 90-104

Getis A, Ord JK (1992). An analysis of spatial association by use of distance statistic. *Geographical Analysis*, 24(3):189-206

Goodchild MF (2004). GIScience, geography, form, and process. *Annals of the Association of American Geographers*, 94(4): 709-714

Griffith DA (2003). *Spatial Autocorrelation and Spatial Filtering: Gaining Understanding Through Theory and Scientific Visualization*. Berlin: Springer

Haggett P, Cliff AD, Frey A (1977). *Locational Analysis in Human Geography*. London: Edward Arnold Ltd.

Hartshorne R (1959). *Perspective on the Nature of Geography*. Chicago: Rand McNally & Company

Hu ZL, Chen YG, Liu T (2018). Three laws of the changes in economic geography. *Economic Geography*, 38(10): 1-4 [In Chinese]

Martin GJ (2005). *All Possible Worlds: A History of Geographical Ideas (4th Revised Edition)*. New York, NY: Oxford University Press

Moran PAP (1948). The interpretation of statistical maps. *Journal of the Royal Statistical Society, Series B*, 37(2): 243-251

Moran PAP (1950). Notes on continuous stochastic phenomena. *Biometrika*, 37: 17-33.

Odland J (1988). *Spatial Autocorrelation*. London: SAGE Publications

Ord JK, Getis A (1995). Local spatial autocorrelation statistics: Distributional issues and an application. *Geographical Analysis*, 27(4): 286-306

Schaefer FK (1953). Exceptionalism in geography: a methodological examination. *Annals of the Association of American Geographers*, 43: 226-249

Tobler W (1970). A computer movie simulating urban growth in the Detroit region. *Economic Geography*, 46(2): 234-240

Tobler W (2004). On the first law of geography: A reply. *Annals of the Association of American Geographers*, 94(2): 304-310

# Appendix: Anselin's derivation

## A1. Basic requirements

In Anselin's seminal paper, he defined two general requirements for a local indicator of spatial association (LISA). The basic requirements are as below: "a. the LISA for each observation gives an indication of the extent of significant spatial clustering of similar values around that observation. b. the sum of LISAs for all observations is proportional to a global indicator of spatial association."

For a statistic $L_i$ based on a variable $y_i$ observed at location $i$, the second requirement of a LISA, may be stated formally as

$$\sum_{i=1}^{n} L_i = \gamma \Lambda, \qquad (A1)$$

where $\Lambda$ is a global indicator of spatial association and $\gamma$ is a scale factor. Unfortunately, based on row-normalized spatial weights matrix, the second requirement cannot be really satisfied in both theoretical derivation and empirical analyses.

The following reasoning process is adapted from Anselin's original paper. For easy understanding, I completely adopt his symbols, but one or more concepts will be changed. For example, row standardization is replaced by row normalization (Table A1).

**Table A1 The symbol system of variables and weights in Anselin's seminal paper**

| Measure | Method | Calculation formula | Property |
|---|---|---|---|
| **Size variable** | Centralization | $z_i = y_i - \bar{y}$, $z_j = y_j - \bar{y}$ | The mean value is 0 |
| | Standardization by z-score | $z_i = \dfrac{y_i - \bar{y}}{\sqrt{m_2}}$, $z_j = \dfrac{y_j - \bar{y}}{\sqrt{m_2}}$ | The mean value is 0 and the standard deviation is 1 |
| **Weight** | Global normalization | $\dfrac{w_{ij}}{S_0} = w_{ij} / \sum_{i=1}^{n}\sum_{j=1}^{n} w_{ij}$ | The sum of weights equals 1 |
| | No normalization | $w_{ij}$ | The sum of weights depends |
| | Row normalization | $\dfrac{w_{ij}}{w_i} = w_{ij} / \sum_{j=1}^{n} w_{ij}$ | The sum of weights equals $n$ |

**Note**: According to Anselin (1995), "the weights $w_{ij}$ may be in row-standardized form, though this is not necessary, and by convention, $w_{ii}=0$." This suggests, both no normalization weights and row normalization weights are acceptable for calculating LISAs. For a variable $y_i$ observed at location $i$, the mean is represented by $\bar{y}$.

## A2. Local Moran's index

A local Moran statistic for an observation $i$ may be defined as

$$I_i = z_i \sum_{j=1}^{n} w_{ij} z_j, \qquad (A2)$$

where $z_i$ or $z_j$ is centralized variable, $w_{ij}$ denotes weights, which may be in row-standardized form or not, though this is not necessary. The sum of local Moran's $I$ is

$$\sum_{i=1}^{n} I_i = \sum_{i=1}^{n} z_i \sum_{j=1}^{n} w_{ij} z_j. \qquad (A3)$$

So for the global indicator, Morn's $I$ is

$$I = \frac{n}{S_0} \frac{\sum_{i=1}^{n}\sum_{j=1}^{n} w_{ij}(x_i - \overline{x})(x_j - \overline{x})}{\sum_{i=1}^{n}(x_i - \overline{x})^2} = \frac{n}{S_0} \frac{\sum_{i=1}^{n}\sum_{j=1}^{n} w_{ij} z_i z_j}{\sum_{i=1}^{n} z_i^2} = \frac{\sum_{i=1}^{n} I_i}{S_0(\frac{1}{n}\sum_{i=1}^{n} z_i^2)} = \frac{1}{S_0 m_2}\sum_{i=1}^{n} I_i, \quad (A4)$$

where

$$S_0 = \sum_{i=1}^{n}\sum_{j=1}^{n} w_{ij} \quad (A5)$$

is the sum of the weights, and

$$m_2 = \frac{1}{n}\sum_{i=1}^{n} z_i^2 \quad (A6)$$

denotes the second moment, a consistent, but not unbiased estimate of the population variance. The factor of proportionality between the sum of the local and the global Moran is

$$\gamma = S_0 m_2. \quad (A7)$$

Please note that equation (A4) is based on global normalization weights. This is the necessary condition to guarantee the validness of equations (A8), (A10), and (A11) given later. Equation (A4) can be expressed as

$$\sum_{i=1}^{n} I_i = S_0 m_2 I = \gamma I. \quad (A8)$$

Note that for a row-standardized spatial weights matrix, $S_0 = n$, so that

$$\gamma = S_0 \frac{1}{n}\sum_{i=1}^{n} z_i^2 = \sum_{i=1}^{n} z_i^2. \quad (A9)$$

And for the standardized variable based on $z$-score, $m_2 = 1$, so that

$$\gamma = S_0 m_2 = S_0 = n. \quad (A10)$$

Therefore, for the row-standardized spatial weights matrix, equation (A8) can be written as

$$\sum_{i=1}^{n} I_i = nI. \quad (A11)$$

The local Moran would then be computed as

$$I_i = \frac{z_i}{m_2}\sum_{j=1}^{n} w_{ij} z_j. \quad (A12)$$

which is actually a local Moran's *I* based on *z*-score of observations $y_i$.

Formally, there seems to be no problem with the above mathematical process. However, in fact, based on row-normalization weights matrix, equations (A4), (A8), and (A11), are not correct. In short, Anselin's LISAs based on row-normalization weights cannot satisfy his second requirement, which specified by equation (A1). Let's see the following mathematical process. The row sum of the weights is

$$w_i = \sum_{j=1}^{n} w_{ij}. \tag{A13}$$

Summing equation (A13) yields

$$\sum_{i=1}^{n} w_i = \sum_{i=1}^{n}\sum_{j=1}^{n} w_{ij} = S_0. \tag{A14}$$

However, the weights based on row normalization is as follows

$$w_{ij}^* = \frac{w_{ij}}{w_i} = w_{ij} / \sum_{j=1}^{n} w_{ij}. \tag{A15}$$

Double summing equation (A15) yields

$$\sum_{i=1}^{n}\sum_{j=1}^{n} w_{ij}^* = \sum_{i=1}^{n}(\frac{1}{w_i}\sum_{j=1}^{n} w_{ij}) = \sum_{i=1}^{n}(1) = n. \tag{A16}$$

No problem can be found equations (A14) and (A16), which is deceiving. The local Moran's indexes based on row-normalization weights is

$$I_i = \frac{z_i}{\sum_{j=1}^{n} w_{ij}} \sum_{j=1}^{n} w_{ij} z_j = \frac{z_i}{w_i} \sum_{j=1}^{n} w_{ij} z_j = z_i \sum_{j=1}^{n} \frac{w_{ij}}{w_i} z_j = z_i \sum_{j=1}^{n} w_{ij}^* z_j. \tag{A17}$$

Summing equation (A17) yields

$$\sum_{i=1}^{n} I_i = \sum_{i=1}^{n} \frac{z_i}{\sum_{j=1}^{n} w_{ij}} \sum_{j=1}^{n} w_{ij} z_j = \sum_{i=1}^{n} z_i \sum_{j=1}^{n} \frac{w_{ij}}{w_i} z_j = \sum_{i=1}^{n} \frac{z_i}{w_i} \sum_{j=1}^{n} w_{ij} z_j \neq \gamma I. \tag{A18}$$

We can never derive a relation similar to equation (A8), which satisfies equation (A1).

## A3. Local Geary's coefficient

Using the same principles as before, a local Geary statistic based on no normalized weights and no standardized variable for each observation I was defined

$$c_i = \sum_{j=1}^{n} w_{ij}(z_i - z_j)^2. \tag{B1}$$

Based on standardized variable, the local Geary coefficient was expressed as

$$c_i = \frac{1}{m_2} \sum_{j=1}^{n} w_{ij}(z_i - z_j)^2. \tag{B2}$$

The notation is the same as before. Without loss of generality, the summation of the $c_i$ over all observations is

$$\sum_{i=1}^{n} c_i = \frac{1}{m_2} \sum_{i=1}^{n}\sum_{j=1}^{n} w_{ij}(z_i - z_j)^2 = n \sum_{i=1}^{n}\sum_{j=1}^{n} w_{ij}(z_i - z_j)^2 / \sum_{i=1}^{n} z_i^2. \tag{B3}$$

In comparison, the global Geary statistic is

$$c = \frac{n-1}{2S_0} \sum_{i=1}^{n} \sum_{j=1}^{n} w_{ij}(z_i - z_j)^2 / \sum_{i=1}^{n} z_i^2 . \tag{B4}$$

Substituting equation (B4) into equation (B3) yields

$$\sum_{i=1}^{n} c_i = \frac{2nS_0}{n-1} c . \tag{B5}$$

Comparing equation (B5) into equation (A1) indicates that the factor of proportionality between the sum of the local and global Geary statistics is

$$\gamma = \frac{2nS_0}{n-1} . \tag{B6}$$

Formally, for row-normalized weights, $S_0=n$; therefore, the proportionality factor is $\gamma=2n^2/(n-1)$.

On the surface, there is no problem with the above mathematical reasoning process. In fact, there is a bug. The row normalized weights was unintentionally replaced by the global normalized weights in the derivation. Based on row normalized weights and standardized variable, the local Geary coefficient is actually as below

$$c_i = \frac{1}{m_2 w_i} \sum_{j=1}^{n} w_{ij}(z_i - z_j)^2 = \frac{1}{m_2} \sum_{j=1}^{n} w_{ij}(z_i - z_j)^2 / \sum_{j=1}^{n} w_{ij} . \tag{B7}$$

From equation (B7) it follows

$$\sum_{i=1}^{n} c_i = n \sum_{i=1}^{n} \frac{1}{w_i} \sum_{j=1}^{n} w_{ij}(z_i - z_j)^2 / \sum_{j=1}^{n} z_i^2 \neq \gamma c . \tag{B8}$$

which cannot satisfy the second requirement defined by Anselin (1995).